\documentclass[onecolumn]{elsart}
\usepackage{amssymb}
\usepackage{amsfonts}
\usepackage{amsmath}
\usepackage{graphicx}
\usepackage{subfigure}
\usepackage{epsfig}

\input{tcilatex}

\begin{document}

\begin{frontmatter}%

\title{Analytical approximation schemes for solving exact
renormalization group equations. II Conformal mappings.}%

\author{C.\ Bervillier}\ead{claude.bervillier@lmpt.univ-tours.fr},\author{ B.\ Boisseau}\ead{bruno.boisseau@lmpt.univ-tours.fr},\author{ H.\ Giacomini}\ead{hector.giacomini@lmpt.univ-tours.fr}%

\address{Laboratoire de Math\'{e}matiques et Physique Th\'{e}orique,\\ UMR 6083 (CNRS),\\
F\'ed\'eration Denis Poisson,\\
Universit\'{e} Fran\c{c}ois Rabelais,\\
Parc de Grandmont, 37200 Tours, France}
%

\begin{abstract}
We present a new efficient analytical approximation scheme to two-point
boundary value problems of ordinary differential equations (ODEs) adapted to
the study of the derivative expansion of the exact renormalization group
equations. It is based on a compactification of the complex plane of the
independent variable using a mapping of an angular sector onto a unit disc.
We explicitly treat, for the scalar field, the local potential
approximations of the Wegner-Houghton equation in the dimension $d=3$ and of
the Wilson-Polchinski equation for some values of $d\in \left] 2,3\right] $.
We then consider, for $d=3$, the coupled ODEs obtained by Morris at the
second order of the derivative expansion. In both cases the fixed points and
the eigenvalues attached to them are estimated. Comparisons of the results
obtained are made with the shooting method and with the other analytical
methods available. The best accuracy is reached with our new method which
presents also the advantage of being very fast. Thus, it is well adapted to
the study of more complicated systems of equations.%
\end{abstract}%

\begin{keyword}
Exact renormalisation group%
\sep
Derivative expansion%
\sep
Critical exponents 
\sep
Two-point boundary value problem 
\PACS
02.30.Hq 
\sep
02.30.Mv 
\sep
02.60.Lj 
\sep
05.10.Cc 
\sep
11.10.Gh 
\sep
64.60.Fr 
\end{keyword}%

\end{frontmatter}%

In a previous article \cite{6188} we presented two analytical approaches for
studying the derivative expansion of the exact renormalization group
equation (ERGE, for reviews and recent pedagogical introductions see \cite%
{Wett,4595}). The two methods, based on the commonly used field expansion,
were shown to be more efficient than the current approaches \cite%
{3478,3642,3553} which implicitly assumed that the simple field expansion
converges in $\left[ 0,\infty \right[ $ \cite{6188} whereas it does not \cite%
{3553}. In the first method, introduced in \cite{6110}, the
infinite-boundary condition is explicitly accounted for via an auxiliary
differential equation (ADE) whereas, in the second, the solutions looked for
are approximated by generalized hypergeometric functions (HFA). Another
method very similar to HFA has almost simultaneously been proposed in \cite%
{6201}, it looks for the solutions under the form of Pad\'{e} approximants.
Both three methods work well (though the ADE method has a wider range of
application) but they are rather heavy to implement (see table \ref{Tab1}).
In the present work we show that a simple conformal mapping onto the unit
disc of a suitably chosen angular sector of the complex plane of the
independent variable, compactifies the originally infinite integration
domain so as to make the series of the field expansion in the new variable
convergent on the whole disc of unit radius.

The paper is organized as follows. In section 1, the principle of the
mapping method is introduced with the example of the Wegner-Houghton RG flow
equation \cite{414} in the local potential approximation (LPA). For the
fixed point solution of this equation in three dimensions, one approximately
knows the location of the closest singularity in the complex plane of the
independent variable \cite{3358}. We show that the best convergence
properties provided by the method correspond to the largest angular sector
compatible with the analyticity of the solution in the original variable.
The calculations of the eigenvalues with the mapping method is shown to be
easy and we provide the best estimates ever obtained up to now of the fixed
point solution and the eigenvalues. In section 2, we consider the
Wilson-Polchinski RG flow equation \cite{440,354} in the LPA. This equation
allows us to illustrate the efficiency of the method for different values of
the dimension $2<d\leq 3$. Again, we provide the best results ever obtained
up to now in three dimensions. We determine the locations of the critical
and multicritical fixed points for $d=3$, $8/3$ and $5/2$ together with the
associated eigenvalues for $d=3$ and $8/3$ with an excellent accuracy. We
pursue the determination of the critical fixed point for values of $d$ very
close to $2$. (At $d=2$, the type of solutions we track disappears.) In
section 3, we look at the second order of the derivative expansion $O\left(
\partial ^{2}\right) $ by considering explicitly the Morris RG equations 
\cite{3357} in three dimensions. These equations are much more difficult to
treat than the previous ones, even in the LPA. Nevertheless, we are able to
determine both the fixed point and the eigenvalues with an accuracy
approaching that obtained with the shooting method$\mathrm{\;}$\textrm{\cite%
{3357,3816}}. An estimate of the subcritical \textquotedblleft
odd-exponent\textquotedblright\ is obtained for the first time $O\left(
\partial ^{2}\right) .$ It, however, does not compare favourably with
existing estimates \cite{319,3437}. We discuss the probable reasons of this
disagreement. Finally we summarize and conclude.

\section{The Wegner-Houghton flow equation in the LPA}

As detailed in \cite{6188}, to which article the reader is invited to refer
for some basical definitions if necessary, the study of the existence of
fixed points and of their stability in the derivative expansion of an ERGE,
amounts to look for regular solutions in $\phi \in \left[ 0,\infty \right[ $
of coupled nonlinear ordinary differential equations (ODE). Here $\phi $ is
the (constant) scalar field. Hence the two boundaries associated to the ODEs
under study are (see \cite{6188}):

\begin{enumerate}
\item $\phi =0$ where the symmetry of interest is imposed to the solution

\item $\phi =\infty $ where a specific behaviour in approaching this point
is imposed to the solution
\end{enumerate}

This problem of solving differential equations with boundary conditions may
be numerically studied using a shooting or a relaxation method but these
methods are not always efficient. Instead, one has often recourse to an
analytical method based on an expansion of the solutions looked for in power
series of $\phi $ about a certain value (the origin $\phi _{0}=0$ \cite{3478}%
, or the minimum of the potential \cite{3642,3553}). One may also consider $%
\phi _{0}$ as an adjustable parameter \cite{3553} with a view to improve the
convergence of the series. Unfortunately these series do not converge in the
whole range $\phi \in \left[ 0,\infty \right[ $ \cite{3553,3358} and the
condition at the second boundary cannot be explicitly imposed.

At this stage, it is useful to consider a concrete example.

\subsection{ Fixed point equation\label{FPE}}

Let us take as a paradigm the Wegner-Houghton equation in the LPA \cite{414}:%
\begin{equation}
\dot{U}=\ln \left[ 1+U^{\prime \prime }\right] +dU+\left( 1-\frac{d}{2}%
\right) \phi \,U^{\prime }\,,  \label{eq:WHflow}
\end{equation}%
in which $d$ is the spatial dimension that we shall set equal to three in
this section, $U$ stands for the potential $U\left( \phi ,t\right) $, with $%
t=-\ln $ $\left( \Lambda /\Lambda _{0}\right) $ where $\Lambda $ is the
\textquotedblleft running\textquotedblright\ momentum scale of reference
compared to a fixed scale $\Lambda _{0}$, and $\dot{U}=\partial U/\partial t$%
, $U^{\prime }=\partial U/\partial \phi $, $U^{\prime \prime }=\partial
^{2}U/\partial \phi ^{2}$.

For the fixed point equation with $d=3$:%
\begin{equation}
\ln \left[ 1+U^{\prime \prime }\right] +3U-\,\frac{\phi }{2}U^{\prime }=0\,,
\label{eq:WHFP}
\end{equation}%
the conditions at the boundaries for the solution $U\left( \phi \right) $
are:%
\begin{eqnarray}
&&U\left( 0\right) =\gamma \,\,,  \label{eq:f*1} \\
&&U^{\prime }\left( 0\right) =0\,,  \label{eq:f*2} \\
&&U\left( \phi \right) \underset{\phi \rightarrow \infty }{\rightarrow }%
G\;\phi ^{6}\,,  \label{eq:f*asy}
\end{eqnarray}%
in which $\gamma $ (or $G$) is the integration constant the value $\gamma
^{\ast }$ (or $G^{\ast }$) of which has to be determined. The connection
parameter $\gamma ^{\ast }$ (or $G^{\ast }$) is often considered as a
substitute to the solution $U^{\ast }\left( \phi \right) $ because this
latter is deduced from the knowledge of $\gamma ^{\ast }$ (or $G^{\ast }$) \
by a simple numerical integration of the ODE.

For convenience, instead of $\gamma $, we shall deal with $r=U^{\prime
\prime }\left( 0\right) $ from which, according to (\ref{eq:WHFP}, \ref%
{eq:f*1}, \ref{eq:f*2}), one deduce $\gamma $ via the relation:%
\begin{equation*}
\gamma =-\frac{1}{3}\ln \left[ 1+r\right] \,.
\end{equation*}

Let us expand $U\left( \phi \right) $ about $\phi =0$ up to a finite order $%
2M$. Accounting for the conditions (\ref{eq:f*1},\ref{eq:f*2}) at the first
boundary (the origin $\phi =0$), it comes:%
\begin{eqnarray}
U_{M}\left( \phi \right) &=&-\frac{1}{3}\ln \left[ 1+r\right] +\frac{1}{2}%
rz+\sum_{n=2}^{M}a_{n}\left( r\right) z^{n}\,,  \label{eq:Tseries1} \\
z &=&\phi ^{2}\,,  \label{eq:z}
\end{eqnarray}%
in which we have introduced $z$ for convenience and the coefficients $%
a_{n}\left( r\right) $ are determined as functions of $r$ so that the fixed
point equation (\ref{eq:WHFP}) is satisfied order by order in powers of $z$.
For example, the two first coefficients read:

\begin{eqnarray*}
a_{2}\left( r\right) &=&-\frac{r(1+r)}{12}\,, \\
a_{3}\left( r\right) &=&\frac{r\,\left( 1+r\right) \,\left( 1+7r\right) }{360%
}\,.
\end{eqnarray*}

It remains to determine the value $r^{\ast }$\ of $r$ so that the condition (%
\ref{eq:f*asy}) at the second boundary is satisfied. To impose this
condition, we need an evaluation for a large value of $z$ of the Taylor
polynomial of degree $M$ introduced in (\ref{eq:Tseries1}). This evaluation
is possible if the radius of convergence $R_{c}$ of the Taylor series in
powers of $z$ is infinite. However, in general, in the derivative expansion
of the ERGE, this is not the case: $R_{c}$ is always finite due to the
presence of singularities in the complex plane of $z$. For the unique
nontrivial fixed point solution of (\ref{eq:WHFP}), knowing the
corresponding $r^{\ast },$ the singularity the closest to the origin has
been numerically located in the complex plane of $z$ at the following point $%
z_{0}$ \cite{3358}:

\begin{eqnarray}
\left\vert z_{0}\right\vert &=&9.7344\,,  \label{eq:z0} \\
\mathrm{Arg}\left( z_{0}\right) &=&0.514\,\pi =\alpha _{c}\,\pi \,,
\label{eq:argz0}
\end{eqnarray}%
which implies that%
\begin{equation}
R_{c}=9.7344\,.  \label{eq:RcvalueWH}
\end{equation}

\begin{figure}[tbp]
\begin{center}
\includegraphics*[width=10cm]{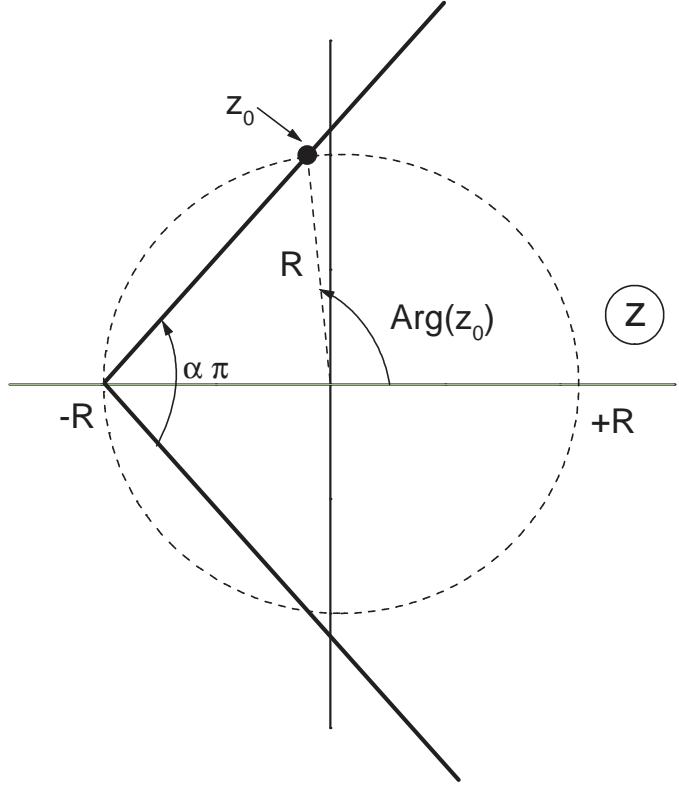}
\end{center}
\caption{When the interior of an angular sector involving the positive real
axis of $z$ defines a region of analyticity for the function $U^{\ast
}\left( \protect\phi \right) $ (with $z=\protect\phi ^{2}$), it may be
mapped onto the interior of a circle of unit radius for the complex variable 
$w$ defined by eq (\protect\ref{eq:Map}). In the present case the
singularity $z_{0}$ of $U^{\ast }\left( \protect\phi \right) $ the closest
to the origin determines the analyticity domain provided the other
singularities lay outside the angular sector containing the origin.}
\label{fig1}
\end{figure}

From the known localization of the singularity the closest to the origin, we
may assume that the solution $U^{\ast }\left( \phi \right) $ we are looking
for is analytical in the angular sector\ which symmetrically straddles the
positive real axis, has its vertex located on the negative real axis at $%
-\left\vert z_{0}\right\vert $ and its angle equal to $\mathrm{Arg}\left(
z_{0}\right) $ (see figure \ref{fig1}). This assumption implies that none of
the eventual supplementary singularities lies inside this sector.

The basis of our approach consists in exploiting the analyticity of the
solution $U^{\ast }$ looked for in the range $z\in \left[ 0,\infty \right[ $%
. Actually, if this solution exists then there is an angular sector
involving the whole positive real axis of the complex plane of $z,$ and
including the origin, in which $U^{\ast }$ is analytic. We thus introduce a
conformal mapping of an angular sector which\ symmetrically straddles this
positive real axis onto the unit disc centered at the origin of the complex
plane of a new variable $w$ defined by:

\begin{equation}
w=\frac{\left( 1+z/R\right) ^{1/\alpha }-1}{\left( 1+z/R\right) ^{1/\alpha
}+1}\,,  \label{eq:Map}
\end{equation}%
in which $R$ and $\alpha $ characterize the position of the vertex and the
angle of an angular sector of the complex plane of $z$ as shown in figure %
\ref{fig1}.

The inverse transformation is:%
\begin{equation}
z=R\left[ \left( \frac{1+w}{1-w}\right) ^{\alpha }-1\right] \,.
\label{eq:MapInverse}
\end{equation}

In particular, if we adjust the angular sector in such a way that the
presently known closest-to-the-origin singularity [see equations (\ref{eq:z0}%
, \ref{eq:argz0})] lies right on one of its edge i.e.: 
\begin{eqnarray}
\alpha &=&\alpha _{c}=0.514\,,  \label{eq:RcWH} \\
R &=&R_{c}\,,  \label{eq:alphaWH}
\end{eqnarray}%
then the resulting Taylor series of the solution $U^{\ast }$ \ expressed in
powers of $w$ should converge on the whole disc $\left\vert w\right\vert <1$.

Consequently, to obtain an auxiliary condition for determining the value $%
r^{\ast },$ we proceed as follows.

We deduce from (\ref{eq:Tseries1}, \ref{eq:MapInverse}) the following Taylor
polynomial:

\begin{equation}
\tilde{U}_{M}\left( w\right) =-\frac{1}{3}\ln \left[ 1+r\right]
+\sum_{n=1}^{M}c_{n}\left( r\right) w^{n}\,,  \label{eq:Tseries2}
\end{equation}%
where $\tilde{U}_{M}\left( w\right) =U_{M}\left[ \phi \left( w\right) \right]
$ with $\phi \left( w\right) $ given by (\ref{eq:z}, \ref{eq:MapInverse}).

Then, from the convergence of the resulting series in powers of $w$, we may:

\begin{enumerate}
\item \label{either}either, impose when $w\rightarrow 1$ (the counterpart of 
$z\rightarrow \infty $), that $\tilde{U}_{M}\left( w\right) $ satisfies the
counterpart of (\ref{eq:f*asy}),

\item \label{or}or, in a less rigorous but often efficient manner,
implicitly exploit the convergence of the Taylor series in powers of $w$ on
the whole unit disc by imposing, as proposed in \cite{1610}, that the $%
M^{th} $ coefficient $c_{M}\left( r\right) $ vanishes (this assumes that the
remaining higher terms of the series do not contribute too much).
\end{enumerate}

Choosing one of these two variants yields an auxiliary condition\ under the
form of a polynomial equation for $r$ the zeros of which are candidates for
the solution $r^{\ast }$ we are looking for.

For the values given by (\ref{eq:RcvalueWH}, \ref{eq:RcWH}, \ref{eq:alphaWH}%
), and the variant (\ref{either}) of the method as example, the real zeros
obtained are distributed almost similarly to those displayed in figure 1 of 
\cite{6188}. An important difference with \cite{6188} however, is that it is
easier to follow the zero of interest as $M$ grows. The estimation we obtain
this way for $r^{\ast }$ is excellent since we get 28 stabilized digits for $%
M=145$:%
\begin{equation}
r^{\ast }=-0.4615337201162071199657576484\,.  \label{eq:rstar}
\end{equation}

It appears that the values of $R$ and $\alpha $ which correspond to the
location of the closest known singularity provide the best convergence for $%
r^{\ast }$. Of course, this situation is particular because, thanks to the
results of \cite{3358}, we know the location of the closest singularity. In
general this is not the case and, as we shall illustrate in section \ref{d=3}%
, the only information practically accessible is a rough estimation of $%
R_{c} $ whereas $\alpha _{c}$ is not known. It is thus justified to try
different values of $R$ and $\alpha $.

Let us define a practical measure of the accuracy of the results as $M$
grows by the number of stabilized digits of $r^{\ast }$:

\begin{equation}
N_{\mathrm{acc}}=-\log \left\vert 1-\frac{r_{M-1}^{\ast }}{r_{M}^{\ast }}%
\right\vert \,.  \label{eq:Nacc}
\end{equation}

\begin{figure}[tbp]
\begin{center}
\includegraphics*[width=10cm]{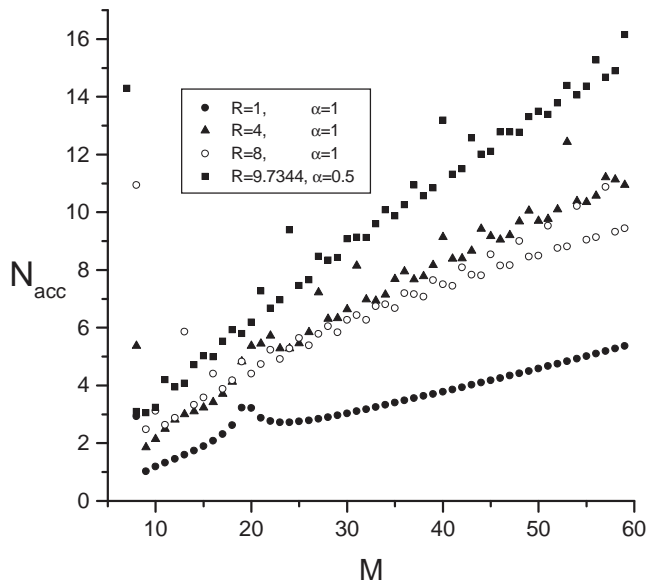}
\end{center}
\caption{Study of the fixed point of the Wegner-Houghton equation in the LPA
($d=3$): approximate number\ $N_{\mathrm{acc}}$\ of stabilized digits on $%
r_{M}^{\ast }$ [see (\protect\ref{eq:Nacc})] as function of the degree $M$
of the Taylor polynomial in $w,$ for different values of the couple $\left\{
R,\protect\alpha \right\} $ which characterizes the conformal mapping (%
\protect\ref{eq:Map})$.$ The most efficient convergence is clearly obtained
for values of $R$ and $\protect\alpha $ corresponding to the location of the
singularity \ the closest to the origin [see (\protect\ref{eq:z0}, \protect
\ref{eq:argz0})].}
\label{fig3}
\end{figure}

Figure \ref{fig3} illustrates well the fact that, if one chooses $R\neq
R_{c} $ and $\alpha \neq \alpha _{c}$, then, in general, one gets a worse
convergence towards the (hopefully true) value of $r^{\ast }$.

But, there are also other values for the couple $\left\{ R,\alpha \right\} $
which provide essentially the same convergence, in particular when $R$ is
chosen greater\footnote{%
Taking into account only the criterium of analyticity, one may choose any
value of $R$ and $\alpha $ such that $R<R_{c}$ and $\alpha <\alpha _{c}$.
But, it often appears that the simultaneous largest possible values of $R$
and $\alpha $ yield the best convergences.} than $R_{c}$. Notice that, in
such a case, $\alpha $ must be smaller than $\alpha _{c}$ otherwise the
singularity located at $z_{0}$ would have an image in the interior of the
unit circle in the complex plane of $w$. These favourable cases
corresponding to $R>R_{c}$ exist because the closest singularity is not
located on the negative real axis of $z$. With the Wilson-Polchinski
equation \cite{440,354}, studied in section (\ref{WilPol}), one encounters
situations where the condition $R\leq R_{c}$ is necessary, indicating that
the closest singularity is right on (or very close to) the negative real
axis of $z$.

\subsubsection{Eigenvalues}

A linearization about the solution $U^{\ast }\left( \phi \right) $ of (\ref%
{eq:WHFP}), using $U\left( \phi ,t\right) =U^{\ast }\left( \phi \right)
+\epsilon \,e^{\lambda t}g\left( \phi \right) $\ with $\epsilon $ a small
parameter, provides the eigenvalue equation:%
\begin{equation*}
\frac{g^{\prime \prime }}{1+U^{\ast \prime \prime }}-\frac{\phi }{2}%
g^{\prime }+\left( 3-\lambda \right) g=0\,,
\end{equation*}%
which may be studied by expanding both $U^{\ast }$ and $g$ in powers of $%
\phi $ (or $z$) as done above in (\ref{eq:Tseries1}, \ref{eq:z}) for the
fixed point.

Accounting for the arbitrariness of the normalization of the eigenfunction,
the conditions at the origin depend on the symmetry of the solution looked
for:

\begin{eqnarray*}
g\left( 0\right) &=&1,\quad g^{\prime }\left( 0\right) =0\,,\qquad \text{%
even case}\,, \\
g\left( 0\right) &=&0,\quad g^{\prime }\left( 0\right) =1\,,\qquad \text{odd
case}\,,
\end{eqnarray*}%
whereas the condition at infinity is:%
\begin{equation}
g\left( \phi \right) \underset{\phi \rightarrow \infty }{\rightarrow }\;\phi
^{6-2\lambda }\,.  \label{eq:eigenasy}
\end{equation}

\begin{figure}[tbp]
\begin{center}
\includegraphics*[width=10cm]{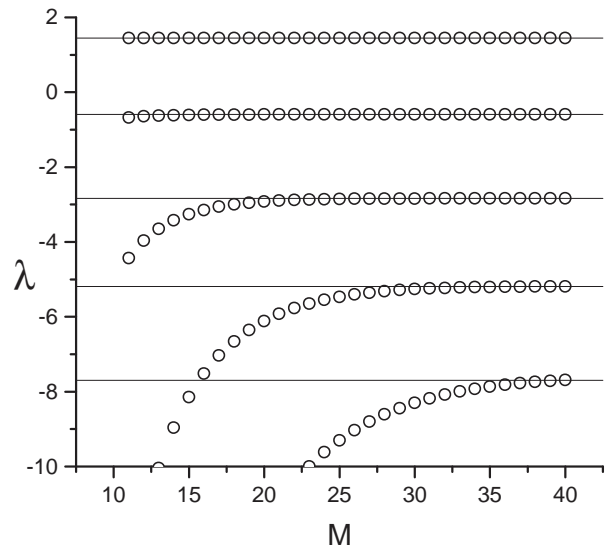}
\end{center}
\caption{Eigenvalue spectrum (even case) of the Wegner-Houghton equation in
the LPA ($d=3$) linearized about the Wilson-Fisher fixed point. The open
circles are the real zeros (as function of the degree of the Taylor
polynomial) of the auxiliary condition provided by the mapping method
(presently using variant (\protect\ref{or})). The horizontal lines represent
the estimates of the eigenvalues (even case). This extremely clear picture
illustrates well the facility for determining the eigenvalues.}
\label{fig4}
\end{figure}

Once $r^{\ast }$ is fixed to the value (\ref{eq:rstar}), the coefficients of
the series for $g$ depend only on $\lambda $ which plays the role of the
connection parameter in the preceding fixed point equation. We perform the
conformal mapping $z\rightarrow w$ defined by (\ref{eq:Map}) with $R$ and $%
\alpha $ fixed to the preceding values, namely $R=9.7344$ and $\alpha =0.514$
and, because the power of $\phi $ in (\ref{eq:eigenasy}) depends on $\lambda 
$ we use the variant (\ref{or}) of the method, although we could use again
the variant (\ref{either}) with the condition $\left( 6-2\lambda \right)
g-\phi g^{\prime }=0$ imposed when $w\rightarrow 1$ on the transformed
series. The real zeros of the resulting polynomial in $\lambda $ are
displayed in figure \ref{fig4} which shows an exceptionally clear
distribution of the spectrum of the eigenvalues.

For simplifying the notation, we generically note $\nu $ the inverse of each
positive eigenvalue $\lambda $ and $\omega $ the opposite of each negative
one in the even case, respectively $\breve{\nu}$ and $\breve{\omega}$ in the
odd case (eventually with an index when there are several eigenvalues of the
same kind for each case, see \cite{6188} for more details):%
\begin{equation*}
\nu =\frac{1}{\lambda }>0\quad ,\quad \omega =-\lambda >0\,.
\end{equation*}

Using an expansion up to $M=104$, we get an estimate of $\nu $ with six
supplementary digits compared to the already excellent determination by Aoki
et al \cite{3553}:%
\begin{equation*}
\nu =0.68945905616213484062727\,.
\end{equation*}

At order $M=40$, the estimates of the eigenvalues for both the even and odd
cases are the following:%
\begin{equation*}
\begin{array}{ll}
\nu =0.68945905616\,, &  \\ 
\omega _{1}=0.59524\,, & \breve{\omega}_{1}=1.691340\,, \\ 
\omega _{2}=2.8385\,, & \breve{\omega}_{2}=3.999\,, \\ 
\omega _{3}=5.19\,, & \breve{\omega}_{3}=6.40\,, \\ 
\omega _{4}=7.7\,, & \breve{\omega}_{4}=9.0\,.%
\end{array}%
\end{equation*}%
The values of the subleading critical exponents are in agreement with the
previous estimates\ \cite{5625}. As in the case of $\nu $, the accuracy of
each of them may be easily improved by considering higher values of $M$.

\section{The Wilson-Polchinski flow equation in the LPA for $2<d<4$\label%
{WilPol}}

The Wilson-Polchinski flow equation \cite{440,354} in the LPA reads:

\begin{equation*}
\dot{U}=U^{\prime \prime }-\left( U^{\prime }\right) ^{2}+\left( 1-\frac{d}{2%
}\right) \phi U^{\prime }+dU\,.
\end{equation*}

By considering this equation we aim at illustrating other aspects of the
method, in particular its efficiency compared to other procedures (when $d=3$%
) but also the way it works when several fixed points exist (when $2<d<3$).
A recent study of this equation for generic $d$ may be found in \cite{6228}
and the estimates that we provide below compare favourably to those obtained
in \cite{5106}.

\subsection{Fixed point equation}

The fixed point ODE reads:%
\begin{equation}
U^{\prime \prime }-\left( U^{\prime }\right) ^{2}+\left( 1-\frac{d}{2}%
\right) \phi U^{\prime }+dU=0\,,  \label{eq:WPFP}
\end{equation}%
the solutions of which we are looking for have the following properties%
\footnote{%
In \cite{6188} the parameter $A$ was noted $b$.}:%
\begin{eqnarray}
&&U\left( 0\right) =k\,,  \label{eq:U*0} \\
&&U^{\prime }\left( 0\right) =0\,,  \label{eq:U'*0} \\
&&U\left( \phi \right) \underset{\phi \rightarrow \infty }{\rightarrow }%
\frac{1}{2}\phi ^{2}-\frac{1}{d}+A\phi ^{2d/(d+2)}+\mathrm{O}\left[ \phi
^{2\left( d-2\right) /(d+2)}\right] \,,  \label{eq:U*asy}
\end{eqnarray}%
where $k$ (or $A$) has to be fixed \ to the value of the connection
parameter $k^{\ast }$ (or $A^{\ast }$).

As in the preceding section, to determine $k^{\ast }$, we consider the
following Taylor polynomial of degree $M$ in $z=\phi ^{2}$:%
\begin{eqnarray*}
U_{M}\left( \phi \right) &=&k+\dsum\limits_{n=0}^{M}b_{n}\left( k\right)
z^{n}\,, \\
z &=&\phi ^{2}\,,
\end{eqnarray*}%
where the coefficients $b_{n}\left( k\right) $ are determined as function of 
$k$ such that the fixed point equation (\ref{eq:WPFP}) is satisfied order by
order in powers of $z$. Then we perform the conformal mapping (\ref{eq:Map}%
). Since we have a priori no information on the analyticity properties of $%
U^{\ast }$ as function of $z,$ we consider $R$ and $\alpha $ as free
parameters. We proceed by trial and error to find their best values: we look
at the variation of the properties of convergence of the estimate $%
k_{M}^{\ast }$ obtained at order $M$ by changing the trial values of $R$ and 
$\alpha $. Let us illustrate this procedure in the case $d=3$ where the
value of $k^{\ast }$ is already known with accuracy \cite{6137}.

\subsubsection{$d=3$\label{d=3}}

We use the variant (\ref{either}) adapted to (\ref{eq:U*asy}) by imposing
the counterpart of $\mathrm{d}U^{\ast }/\mathrm{d}z\rightarrow 1/2$ when $%
w\rightarrow 1$. Setting arbitrarily $R=1$ and $\alpha =1$, we easily
distinguish, among all the real zeros of the polynomial auxiliary condition
obtained at order $M$, a generic zero which presumably will converge to $%
k^{\ast }$ when $M\rightarrow \infty $. The observed convergence is not
excellent\ yet, but it is sufficient to allow a rough estimate of the radius
of convergence $R_{c}$ for the expected $k^{\ast }$. Following the
d'Alembert or Cauchy rules, this estimate may be obtained by looking at:%
\begin{equation*}
R_{1,n}=\left\vert \frac{b_{n}}{b_{n+1}}\right\vert \,,
\end{equation*}%
or:%
\begin{equation*}
R_{2,n}=\left\vert b_{n}\right\vert ^{-1/n}\,,
\end{equation*}%
for a given value of $k$ and large values of $n$.

In general, we have proceeded that way, by trial and error, to estimate $%
R_{c}$. In the present case, however, the radius of convergence of the fixed
point series is known to be \cite{6188}:%
\begin{equation*}
R_{c}=5.72167\,.
\end{equation*}

\begin{figure}[tbp]
\begin{center}
\includegraphics*[width=10cm]{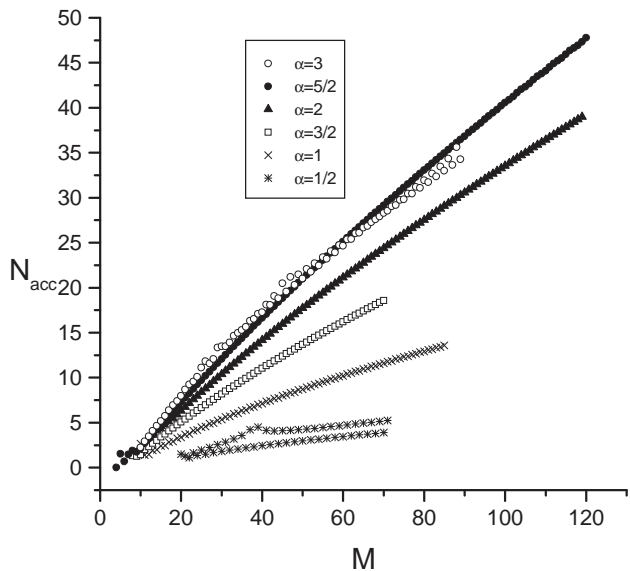}
\end{center}
\caption{Study of the fixed point of the Wilson-Polchinski equation in the
LPA for $d=3$: approximate number\ $N_{\mathrm{acc}}$\ of stabilized digits
on $k_{M}^{\ast }$ [similar to (\protect\ref{eq:Nacc})] as function of the
degree $M$ of the Taylor polynomial in $w$, for different values of $\protect%
\alpha $ and for $R=5.72167.$}
\label{fig5}
\end{figure}

By setting straight out $R$ to this value (to save time), we get a better
accuracy on the estimation of $k^{\ast }$ than with $R=1$, and on varying $%
\alpha $, we observe that the accuracy is better and better when $\alpha $
grows. Finally, as shown in figure \ref{fig5} the best result is obtained
with $\alpha =5/2$. This high value of $\alpha $ suggests that the
singularity is located on the negative real axis and is unique. Indeed when
we try values of $R$ larger than $R_{c}$, the zero of interest disappears
contrary to the preceding Wegner-Houghton case for which the singularity the
closest to the origin lies outside the negative real axis. The fact that we
can choose $\alpha >2$ (when $R=R_{c}$) without spoiling the property of
convergence of the mapping method, suggests that this singularity is unique.

We finally get an accuracy of $49$ digits for $M=120$:%
\begin{equation}
k^{\ast }=0.07619940081234064145788536913234906280801814336214\pm 6\times
10^{-50}\,.  \label{eq:kstarlong}
\end{equation}

\begin{table}[tbp] \centering%
\begin{tabular}{|c|c|c|}
\hline\hline
method & $k^{\ast }$ & time \\ \hline
\multicolumn{1}{|l|}{ADE} & \multicolumn{1}{|l|}{0.076199400812365} & 
\multicolumn{1}{|r|}{1523.84} \\ 
\multicolumn{1}{|l|}{Pad\'{e}} & \multicolumn{1}{|l|}{0.07619940081205} & 
\multicolumn{1}{|r|}{1364.73} \\ 
\multicolumn{1}{|l|}{HFA} & 0.076199400812340 & \multicolumn{1}{|r|}{138.58}
\\ 
mapping & \multicolumn{1}{|l|}{0.0761994008160} & \multicolumn{1}{|r|}{2.00}
\\ \hline
\end{tabular}%
\caption{Comparison between estimates of the connection parameter $k^*$, of the Wilson-Polchinski RG equation
 in the LPA ($d=3$),
obtained using different efficient analytical methods at order $M=25$ of the Taylor polynomial [compare to (\ref{eq:kstarlong})]. 
The ``time'' given in the third column is a CPU time (in seconds) corresponding to the
calculation, on the same computer, using each method.}\label{Tab1}%
\end{table}%

Table \ref{Tab1} shows a comparison of efficiencies at $M=25$ between the
four methods ADE, HFA, Pad\'{e} and the present mapping method. This latter
method appears to be, by far, the most efficient.

The mapping method provides also a global explicit representation of the
solution via the Taylor polynomial in powers of $w$ re-expressed in terms of 
$\phi $ (the HFA and Pad\'{e} methods give also a global representation of
the solution but the procedure is much more heavy). From this global
representation, knowing $k^{\ast }$, it is not difficult to estimate the
value of the connection parameter $A^{\ast }$ which appears in (\ref%
{eq:U*asy}). With three more terms in the asymptotic behaviour of $U^{\ast }$
than in (\ref{eq:U*asy}), we have been able to obtain the following result:%
\begin{equation*}
A^{\ast }\simeq -2.3184\,,
\end{equation*}%
which is very close to the value we have determined with the shooting method 
\cite{6188}:%
\begin{equation*}
A^{\ast }\simeq -2.3183\,.
\end{equation*}

Once the fixed point has been determined, the eigenvalues are looked for
with the same couple ($R=R_{c},$ $\alpha =5/2)$ and, as in the preceding
case, we easily get the eigenvalues with high accuracy. We may thus give the
best estimates ever obtained up to now (see \cite{6188} for the definitions
of both the eigenvalue equation and the exponents, and compare the following
values with the estimates obtained in \cite{6137}):

\begin{itemize}
\item for $M=75$ in the even case ($d=3$):%
\begin{eqnarray*}
\nu &=&0.649561773880648017614299724015827\pm 2\times 10^{-33}\,, \\
\omega _{1} &=&0.6557459391933387407836879749684\pm 2\times 10^{-31}\,, \\
\omega _{2} &=&3.180006512059167532314140242\,, \\
\omega _{3} &=&5.912230612747701026351105\,, \\
\omega _{4} &=&8.796092825413903643907\,, \\
\omega _{5} &=&11.798087658336857239\,.
\end{eqnarray*}

\item for $M=69$ in the odd case ($d=3$):%
\begin{eqnarray*}
\breve{\omega}_{1} &=&1.8867038380914203710417873172\pm 5.3\times 10^{-28}\,,
\\
\breve{\omega}_{2} &=&4.524390733670772780436353\,, \\
\breve{\omega}_{3} &=&7.3376506433543135387526\,, \\
\breve{\omega}_{4} &=&10.2839007240259581722\,, \\
\breve{\omega}_{5} &=&13.3361699643459431\,.
\end{eqnarray*}
\end{itemize}

\subsubsection{$d=8/3$}

A new non-trivial fixed point emerges from the Gaussian fixed point when $d$
takes on a value below each threshold \cite{2080}:%
\begin{equation*}
d=2\frac{p+1}{p}\,,
\end{equation*}%
where $p$ takes on the integer values $1,2,3,\cdots ,\infty $. Each one of
these dimensions corresponds to a monomial in powers of $\phi $, in the
formally expanded potential, becoming relevant with respect to the Gaussian
fixed point. Hence for $3\leq d<4$, there is only one nontrivial fixed
point: the Wilson-Fisher fixed point \cite{439} which controls the behaviour
of any Ising-like critical point. Below $d=3$ appears a new fixed point
which controls the tri-critical behaviour. Down to $d=8/3$ there are two
fixed points.

To determine these two fixed points, we proceed similarly to the previous
case $d=3$ by trial and error in order to determine the best values for $R$
and $\alpha $. It appears that we easily determine the critical fixed point
and get a rather good convergence for $R=3.5$ and $\alpha =2$. We finally
obtain with $M=120$ the following estimation:%
\begin{equation*}
k_{c}^{\ast }=0.16736641293800245119399231271370187\pm 3\times 10^{-35}\,.
\end{equation*}

It is worth indicating that the radius of convergence of the series for this
value of $k^{\ast }$ is:%
\begin{equation*}
R_{c}=3.539\,.
\end{equation*}

An estimation of this connection parameter by a shooting method gives the
following result:%
\begin{equation*}
k_{c}^{\ast }=0.16736641293\,.
\end{equation*}

For the same values of $R$ and $\alpha $, and the variant (\ref{or}), we get
the following estimates for the (critical) eigenvalues (without trying to
optimize the accuracy):

\begin{itemize}
\item in the even case for $M=50$:%
\begin{eqnarray*}
\nu &=&0.76520486063609135\pm 1.8\times 10^{-16}\,, \\
\omega _{1} &=&0.81489673946448431495\pm 7.4\times 10^{-15}\,, \\
\omega _{2} &=&3.32048207983\,, \\
\omega _{3} &=&6.045184748\,, \\
\omega _{4} &=&8.9242446\,, \\
\omega _{5} &=&11.92109\,,
\end{eqnarray*}

\item in the odd case for $M=60$:%
\begin{eqnarray*}
\breve{\omega}_{1} &=&2.032987010721739017\pm 5\times 10^{-18}\,, \\
\breve{\omega}_{2} &=&4.66036178907299\,, \\
\breve{\omega}_{3} &=&7.46810111086\,, \\
\breve{\omega}_{4} &=&10.409630969\,, \\
\breve{\omega}_{5} &=&13.4559703\,.
\end{eqnarray*}
\end{itemize}

The tri-critical fixed point is more difficult to determine than the
critical one because a clear identification of the correct zero is only
possible above $M\simeq 20,$ \ a value which is greater than that observed
in the case of the critical fixed point. Nevertheless, we get a good
estimate for $R=19.5$ et $\alpha =0.47$ (variant (\ref{or}), and $M=120$):%
\begin{equation*}
k_{t}^{\ast }=-0.0152088617493\pm 9.2\times 10^{-13}\,,
\end{equation*}%
whereas by the shooting method we get:%
\begin{equation*}
k_{t}^{\ast }=-0.015208861395\,,
\end{equation*}

The radius of convergence of the series in this case is:%
\begin{equation*}
R_{c}\simeq 19.4\,.
\end{equation*}

For the same values of $R$ and $\alpha $, and the variant \ref{or}, we get
the following estimates for the (tri-critical) eigenvalues (without trying
to optimize the accuracy):

\begin{itemize}
\item in the even case:
\end{itemize}

\begin{eqnarray*}
\nu _{1} &=&0.501489558\,, \\
\nu _{2} &=&1.0479506\,, \\
\omega _{1} &=&0.343226\,, \\
\omega _{2} &=&1.83594\,, \\
\omega _{3} &=&3.484\,, \\
\omega _{4} &=&5.26\,, \\
\omega _{5} &=&7.2\,,
\end{eqnarray*}

\begin{itemize}
\item in the odd case:%
\begin{eqnarray*}
\breve{\nu}_{1} &=&0.659916\,, \\
\breve{\omega}_{1} &=&1.0682\,, \\
\breve{\omega}_{2} &=&2.643\,, \\
\breve{\omega}_{3} &=&4.36\,, \\
\breve{\omega}_{4} &=&6.27\,, \\
\breve{\omega}_{5} &=&8.4\,.
\end{eqnarray*}
\end{itemize}

As expected, compared to the critical fixed point, the tri-critical fixed
point has one additional positive even eigenvalue ($\lambda =1/\nu _{2}$),
and also one additional positive odd eigenvalue ($\lambda =1/\breve{\nu}_{1}$%
).

\subsubsection{Lower dimensions}

The calculations are more and more difficult as $d$ decreases. This is
because the radius of convergence $R_{c}\left( d\right) $ for a given kind
of fixed point (critical or multi-critical) also decreases. Consequently, in
approaching $d=2,$ the determination of the critical fixed point itself is
made more and more difficult.\ In fact, at $d=2$, the regular solution we
are searching (which satisfies the condition (\ref{eq:U*asy}) when $\phi
\rightarrow \infty $) disappears. Indeed, the same reasoning as that
followed by Morris in \cite{4424}, shows that for $d=2$ the only solutions\
to the fixed point equation in the LPA are either singular at finite $\phi $
or periodic (see also \cite{4222}).

For the sake of shortness, we have simply tested that we can determine the
critical fixed point (we have not considered the eigenvalues despite the
fact that they are easy to determine) down to dimensions very close to $d=2$%
. Table \ref{Tab3} shows some of the results for $k^{\ast }$. It shows also
that the radius of convergence of the original series in powers of $\phi $
decreases when the location of the minimum of the critical fixed point
potential increases. Consequently, because this property is presumably
general in the derivative expansion, the method of \cite{3642,3553} based on
an expansion about this minimum is presumably doomed to failure before
reaching $d=2$.

\begin{table}[tbp] \centering%
\begin{tabular}{ccccccccccc}
\hline\hline
$d$ &  & $k^{\ast }$ &  & $R$ &  & $\alpha $ &  & $\phi _{0}$ &  & $%
r_{c}(\phi )$ \\ \hline
\multicolumn{1}{l}{$3$} &  & \multicolumn{1}{l}{$0.07619940081234$} &  & 
\multicolumn{1}{l}{$5.7$} &  & \multicolumn{1}{l}{$2.5$} &  & $1.905202$ & 
& $2.39$ \\ 
\multicolumn{1}{l}{$8/3$} &  & \multicolumn{1}{l}{$0.16736641293800$} &  & 
\multicolumn{1}{l}{$3.5$} &  & \multicolumn{1}{l}{$2$} &  & $1.991622$ &  & $%
1.87$ \\ 
\multicolumn{1}{l}{$5/2$} &  & \multicolumn{1}{l}{$0.25299549297691$} &  & 
\multicolumn{1}{l}{$2.7$} &  & \multicolumn{1}{l}{$2$} &  & $2.040021$ &  & $%
1.64$ \\ 
\multicolumn{1}{l}{$12/5$} &  & \multicolumn{1}{l}{$0.33045938745626$} &  & 
\multicolumn{1}{l}{$2$} &  & \multicolumn{1}{l}{$2$} &  & $2.077695$ &  & $%
1.51$ \\ 
\multicolumn{1}{l}{$7/3$} &  & \multicolumn{1}{l}{$0.40022546744981$} &  & 
\multicolumn{1}{l}{$1.5$} &  & \multicolumn{1}{l}{$2$} &  & $2.108868$ &  & $%
1.41$ \\ 
\multicolumn{1}{l}{$21/10$} &  & \multicolumn{1}{l}{$0.96514877904597$} &  & 
\multicolumn{1}{l}{$0.75$} &  & \multicolumn{1}{l}{$2$} &  & $2.214026$ &  & 
$1$ \\ 
\multicolumn{1}{l}{$202/100$} &  & \multicolumn{1}{l}{$1.8411524065675$} & 
& \multicolumn{1}{l}{$0.55$} &  & \multicolumn{1}{l}{$2$} &  & $2.370782$ & 
& $0.77$ \\ \hline
\end{tabular}%
\caption{Determination of the connection parameter $k^*$ of the critical-fixed-point solutions of (\ref{eq:WPFP})
 for decreasing values of the spatial dimension $d$. $\phi_0$ stands for the location of the minimum of the fixed point potential 
($U^{*\prime}(\phi_0)=0$) as determined from the present mapping method. $r_{c}(\phi)=\sqrt{R_{c}(z)}$ is the radius of 
convergence of the original series in powers of $\phi$. Notice that $r_{c}(\phi)$ decreases when the
location of the minimum of $U^*$ increases and becomes smaller than $|\phi_0|$. $R$ and $\alpha$ 
are the parameters of the conformal mapping defined in (\ref{eq:Map}).
}\label{Tab3}%
\end{table}%

Regarding the fixed points of higher criticality, we have not tried to
determine them below $d=5/2.$ For this value of $d$ we get the locations of
the fixed points displayed in table \ref{Tab2}.

\begin{table}[tbp] \centering%
\begin{tabular}{|l|l|c|c|c|}
\hline\hline
type of fixed point & $k^{\ast }\left\{ 
\begin{array}{c}
\mathrm{mapping} \\ 
\mathrm{shooting}%
\end{array}%
\right. $ & $\min $($M)$ & $\left\{ 
\begin{array}{c}
R \\ 
R_{c}%
\end{array}%
\right. $ & $\alpha $ \\ \hline
& \multicolumn{1}{|c|}{} &  &  &  \\ 
critical & $\left\{ 
\begin{array}{l}
0.2529954929769066647431 \\ 
0.252995492980%
\end{array}%
\right. $ & $15$ & \multicolumn{1}{|r|}{$\left\{ 
\begin{array}{c}
2.71 \\ 
2.71%
\end{array}%
\right. $} & 2 \\ 
& \multicolumn{1}{|c|}{} &  &  &  \\ 
\multicolumn{1}{|l|}{tri-critical} & \multicolumn{1}{|l|}{$\left\{ 
\begin{array}{l}
-0.043027 \\ 
\multicolumn{1}{c}{-0.043026994781}%
\end{array}%
\right. $} & $50$ & \multicolumn{1}{|r|}{$\left\{ 
\begin{array}{r}
9 \\ 
\sim 13%
\end{array}%
\right. $} & 0.6 \\ 
& \multicolumn{1}{|c|}{} &  &  &  \\ 
quadri-critical & $\left\{ 
\begin{array}{l}
0.0030001 \\ 
\multicolumn{1}{c}{0.003000121150}%
\end{array}%
\right. $ & $40$ & \multicolumn{1}{|r|}{$\left\{ 
\begin{array}{c}
30 \\ 
\sim 27%
\end{array}%
\right. $} & 0.5 \\ 
& \multicolumn{1}{|c|}{} &  &  &  \\ \hline
\end{tabular}%
\caption{Determination of the values of the connection parameter $k^*$  for the three non-trivial fixed points of (\ref{eq:WPFP}) when $d=5/2$. min($M$) indicates
the minimal value of the degree $M$ of the Taylor polynomial at which, in the mapping method, we can
unambiguously identify a zero that one may easily track when $M$ grows. The estimates obtained by using the shooting method are shown on the second line within a bracket. $R$ and $\alpha$
 are the parameters of the conformal mapping (\ref{eq:Map}), the $R_c$ are estimated values of the radius of convergence of the solution
in terms of the original independent variable $z=\phi^2$.}\label{Tab2}%
\end{table}%

\begin{figure}[tbp]
\begin{center}
\includegraphics*[width=10cm]{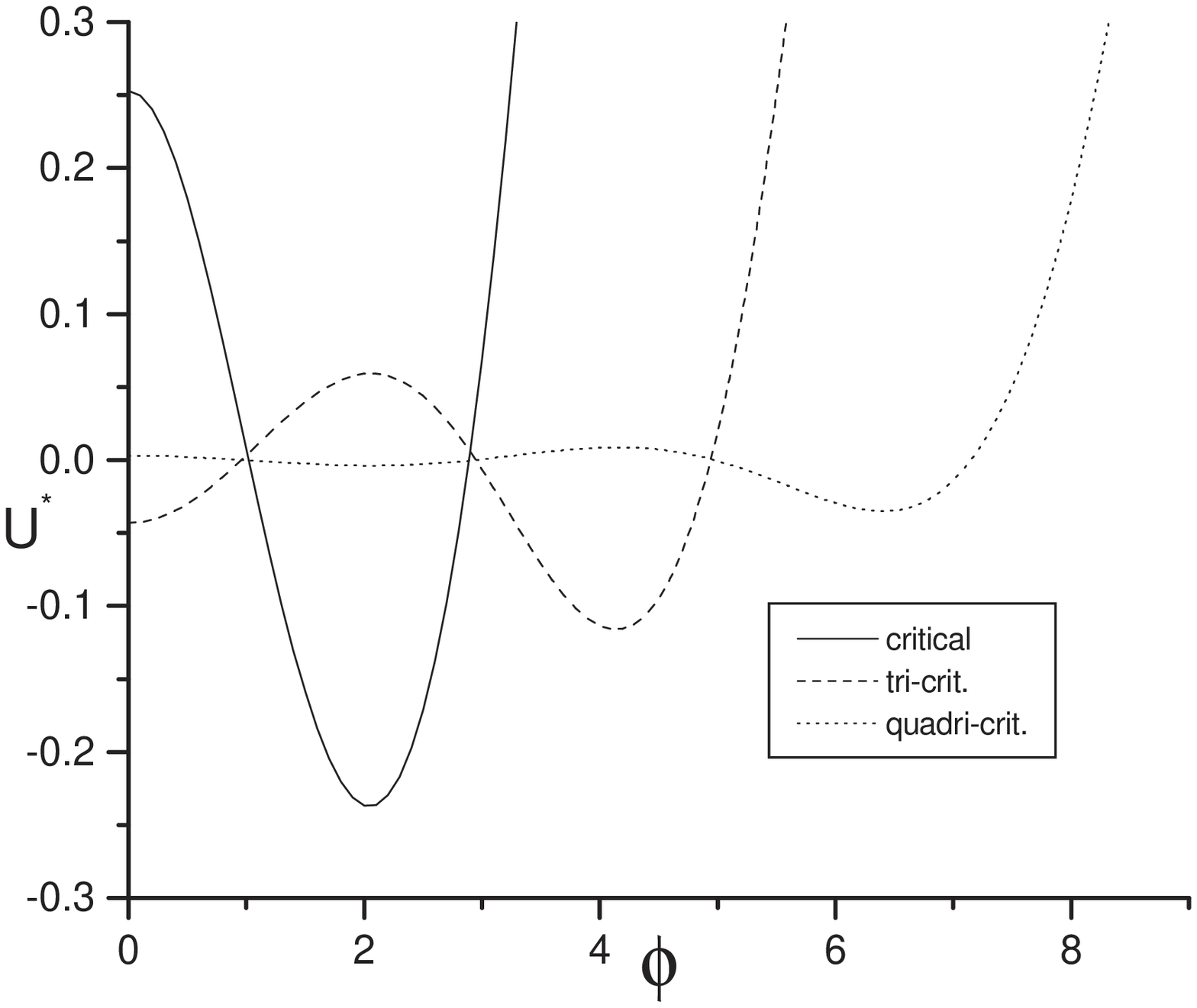}
\end{center}
\caption{The mapping method provides a precise global explicit information
on the solutions looked for. Here are shown the $n$-well potentials as
obtained from the Taylor polynomials in $w\left( \protect\phi \right) $ of
the three fixed-point solutions found for the Wilson-Polchinski RG equation
in the LPA when $d=5/2$.}
\label{fig6}
\end{figure}

We show in figure \ref{fig6} the global solutions we get from the Taylor
polynomials in $w$ for the fixed points when $d=5/2$. They have the right $n$%
-well potential form with $n$ growing with the number of directions of
instability of the fixed point in agreement with \cite{2080}.

\section{The Morris equations (LPA and second order of the derivative
expansion)}

At second order of the derivative expansion [$O\left( \partial ^{2}\right) $
in short], in addition to the potential $U\left( \phi ,t\right) $, a new
function appears: the coefficient $Z\left( \phi ,t\right) $\ of the kinetic
term of the action $S\left[ \phi \right] $. The flow equations for $U$ and $%
Z $ are coupled and the reparameterization invariance of the ERGE \cite{2835}
is broken \cite{4421,3357} together with the so-called scheme invariance
(invariance with respect to a change of the cut-off function which
introduces the running momentum scale of reference $\Lambda $). This is, at
least, the case for the Wilson-Polchinski ERGE. Actually the two invariances
are not independent \cite{4753} and if, instead of the action $S\left[ \phi %
\right] ,$ one considers the Legendre transformed or effective action $%
\Gamma \left[ \varphi \right] $, then the breaking of the invariances may be
reduced to a single one \cite{Wett}. One may even go further and find an
appropriate cut-off function so that the reparameterization invariance is
effectively restored. Morris \cite{3357} has thus obtained the following two
equations (written below for $d=3$ and keeping the original notations%
\footnote{%
Going from $S\left[ \phi \right] $ to $\Gamma \left[ \varphi \right] $
changes the couple $\left\{ U,Z\right\} $ into $\left\{ V,K\right\} .$}): 
\begin{eqnarray}
\dot{V} &=&-\frac{1-\eta /4}{\sqrt{K}\sqrt{V^{\prime \prime }+2\sqrt{K}}}+3V-%
\frac{1}{2}\left( 1+\eta \right) xV^{\prime }\,,  \label{eq:Morris1} \\
\dot{K} &=&\left( 1-\eta /4\right) \left\{ \frac{1}{48}\frac{24KK^{\prime
\prime }-19\left( K^{\prime }\right) ^{2}}{K^{3/2}\left( V^{\prime \prime }+2%
\sqrt{K}\right) ^{3/2}}\right.  \notag \\
&&-\frac{1}{48}\frac{58V^{\prime \prime \prime }K^{\prime }\sqrt{K}+57\left(
K^{\prime }\right) ^{2}+\left( V^{\prime \prime \prime }\right) ^{2}K}{%
K\left( V^{\prime \prime }+2\sqrt{K}\right) ^{5/2}}  \notag \\
&&\left. +\frac{5}{12}\frac{\left( V^{\prime \prime \prime }\right)
^{2}K+2V^{\prime \prime \prime }K^{\prime }\sqrt{K}+\left( K^{\prime
}\right) ^{2}}{\sqrt{K}\left( V^{\prime \prime }+2\sqrt{K}\right) ^{7/2}}%
\right\}  \notag \\
&&-\frac{1}{2}\left( 1+\eta \right) xK^{\prime }-\eta K\,,
\label{eq:Morris2}
\end{eqnarray}%
where $x$ stands for $\varphi $ and $\eta $ is a parameter to be adjusted in
order to find a fixed point.

Let us first consider the determination of the Wilson-Fisher fixed point.

The fixed point equations correspond to:%
\begin{eqnarray}
\dot{V} &=&0\,,  \label{eq:MorrisFP1} \\
\dot{K} &=&0\,,  \label{eq:MorrisFP2}
\end{eqnarray}%
where $\dot{V}$ and $\dot{K}$ stand for the rhs of (\ref{eq:Morris1}, \ref%
{eq:Morris2}).

We look for solutions under the form of even functions of $x$ which must
satisfy the following conditions:%
\begin{eqnarray}
&&V\left( 0\right) =k\,,  \label{eq:MorrisOri1} \\
&&V^{\prime }\left( 0\right) =0\,,  \label{eq:MorrisOri2} \\
&&K\left( 0\right) =K_{0}\,,  \label{eq:MorrisOri3} \\
&&K^{\prime }\left( 0\right) =0\,,  \label{eq:MorrisOri4} \\
&&V\left( x\right) \underset{x\rightarrow \infty }{\simeq }G_{1}x^{\frac{6}{%
1+\eta }}\,,  \label{eq:MorrisAsy1} \\
&&K\left( x\right) \underset{x\rightarrow \infty }{\simeq }G_{2}x^{-\frac{%
2\eta }{1+\eta }}\,.  \label{eq:MorrisAsy2}
\end{eqnarray}%
and which have no singularity in the whole range $x\in \left[ 0,\infty %
\right[ $.

The two parameters $k$ and $\eta $ (or $G_{1}$ and $G_{2}$) have to be
adjusted to the two connection parameters $k^{\ast }$ and $\eta ^{\ast }$
respectively (or $G_{1}^{\ast }$ and $G_{2}^{\ast }$) which define the fixed
point solution of interest\footnote{%
The parameters $k^{\ast }$ and $\eta ^{\ast }$ are related to the parameter $%
\sigma $ and $\eta $ used in \cite{3357} by $\eta ^{\ast }=\eta $ and $%
k^{\ast }=\frac{\left( 1-\eta /4\right) }{3\sqrt{2+\sigma }}.$}. By virtue
of the reparameterization invariance, $K_{0}$ may be set arbitrarily equal
to one without changing the value $\eta ^{\ast }$ (and of the eigenvalues
associated to the linearization of the flow equation about the fixed point).

As in the case of the LPA, we express the two functions as Taylor
polynomials of degree $M$ in $z=x^{2}$:%
\begin{eqnarray*}
V_{M}\left( x\right) &=&k+\dsum\limits_{n=0}^{M}b_{n}^{\left( 1\right)
}\left( k,\eta \right) z^{n}\,, \\
K_{M}\left( x\right) &=&1+\dsum\limits_{n=0}^{M}b_{n}^{\left( 2\right)
}\left( k,\eta \right) z^{n}\,, \\
z &=&x^{2}\,,
\end{eqnarray*}%
in which the coefficients $b_{n}^{\left( 1\right) }\left( k,\eta \right) $
and $b_{n}^{\left( 2\right) }\left( k,\eta \right) $ are determined so as to
satisfy the fixed point equations (\ref{eq:MorrisFP1}, \ref{eq:MorrisFP2}).
To determine the values $k^{\ast }$ and $\eta ^{\ast }$ using these two
polynomials, we need two auxiliary conditions that we deduce from the
conditions (\ref{eq:MorrisAsy1}, \ref{eq:MorrisAsy2}) and the analyticity of
the two functions $V\left( x\right) $ and $K\left( x\right) $ in an angular
sector of the complex plane of the independent variable $z=x^{2}$.

We thus introduce the same conformal mapping $z\rightarrow w$ as defined by (%
\ref{eq:Map}) and we adopt one of the two variants defined in section \ref%
{FPE} to get the two auxiliary conditions either by directly imposing on the
Taylor polynomials the counterpart of the asymptotic conditions (\ref%
{eq:MorrisAsy1}, \ref{eq:MorrisAsy2}) when $w\rightarrow 1$ or, less
rigorously, by assuming the simultaneous vanishing of the two coefficients $%
b_{M}^{\left( 1\right) }\left( k,\eta \right) $ and $b_{M}^{\left( 2\right)
}\left( k,\eta \right) $.

\subsection{LPA}

Let us first study the LPA of (\ref{eq:Morris1}, \ref{eq:Morris2}) which
then reduce to the following unique equation:%
\begin{equation*}
\dot{V}=-\frac{1}{\sqrt{V^{\prime \prime }+2}}+3V-\frac{x}{2}V^{\prime }\,.
\end{equation*}

The numerical study of this flow equation is much more complicated than that
of the Wegner-Houghton or Wilson-Polchinski cases. The largest degree of the
Taylor polynomial reasonnably accessible is, in general, smaller than in
these preceding cases. Nevertheless, using the mapping method, we have
correctly estimated the fixed point location and the eigenvalues with a
better accuracy than the results available in the litterature \textrm{\cite%
{3357,3816}} (obtained by the shooting method, see table \ref{Tab4}). For a
better comparison, we have improved those latter estimates (see table \ref%
{Tab4}).

\begin{table}[tbp] \centering%
\begin{tabular}{cc|c}
\hline\hline
& LPA & $O\left( \partial ^{2}\right) $ \\ 
& $\left\{ 
\begin{array}{l}
\mathrm{mapping\;method} \\ 
\mathrm{shooting\;(this\;work)} \\ 
\mathrm{shooting\;(previous)}%
\end{array}%
\right. $ & $\left\{ 
\begin{array}{l}
\mathrm{mapping\;method} \\ 
\mathrm{shooting\;(this\;work)} \\ 
\mathrm{shooting\;(previous)}%
\end{array}%
\right. $ \\ \hline
&  &  \\ 
$k^{\ast }$ & \multicolumn{1}{l|}{$\left\{ 
\begin{array}{l}
0.2753644064810282 \\ 
0.2753644064810124 \\ 
0.275364406%
\end{array}%
\right. $} & \multicolumn{1}{|l}{$\left\{ 
\begin{array}{l}
0.258216 \\ 
0.258214391 \\ 
0.258\,2144%
\end{array}%
\right. $} \\ 
&  &  \\ 
$\eta ^{\ast }$ & $0$ & \multicolumn{1}{|l}{$\left\{ 
\begin{array}{l}
0.053941 \\ 
0.0539320839 \\ 
0.05393208%
\end{array}%
\right. $} \\ 
&  &  \\ 
$\nu $ & \multicolumn{1}{l|}{$\left\{ 
\begin{array}{l}
0.660389431 \\ 
0.660389431331 \\ 
0.660389%
\end{array}%
\right. $} & \multicolumn{1}{|l}{$\left\{ 
\begin{array}{l}
0.618063 \\ 
0.61806 \\ 
0.6181%
\end{array}%
\right. $} \\ 
&  &  \\ 
$\omega _{1}$ & \multicolumn{1}{l|}{$\left\{ 
\begin{array}{l}
0.6285575 \\ 
\multicolumn{1}{c}{0.6285575035} \\ 
0.6285%
\end{array}%
\right. $} & \multicolumn{1}{|l}{$\left\{ 
\begin{array}{l}
0.8964 \\ 
0.897227 \\ 
0.8975%
\end{array}%
\right. $} \\ 
&  &  \\ 
$\omega _{2}$ & \multicolumn{1}{l|}{$\left\{ 
\begin{array}{l}
3.04801 \\ 
\multicolumn{1}{c}{3.048005033} \\ 
\multicolumn{1}{c}{\mathrm{-}}%
\end{array}%
\right. $} & \multicolumn{1}{|l}{$\left\{ 
\begin{array}{l}
1.71 \\ 
1.7 \\ 
\multicolumn{1}{c}{\mathrm{-}}%
\end{array}%
\right. $} \\ 
&  &  \\ 
$\breve{\omega}_{1}$ & \multicolumn{1}{l|}{$\left\{ 
\begin{array}{c}
1.8124863608 \\ 
\multicolumn{1}{l}{1.812486361} \\ 
\mathrm{-}%
\end{array}%
\right. $} & \multicolumn{1}{|l}{$\left\{ 
\begin{array}{l}
0.86562 \\ 
0.865569 \\ 
\multicolumn{1}{c}{\mathrm{-}}%
\end{array}%
\right. $} \\ 
&  &  \\ 
$\breve{\omega}_{2}$ & \multicolumn{1}{l|}{$\left\{ 
\begin{array}{l}
4.32251050 \\ 
\multicolumn{1}{c}{4.3225104975} \\ 
\multicolumn{1}{c}{\mathrm{-}}%
\end{array}%
\right. $} & \multicolumn{1}{|l}{$\left\{ 
\begin{array}{l}
2.8 \\ 
\multicolumn{1}{c}{\mathrm{-}} \\ 
\multicolumn{1}{c}{\mathrm{-}}%
\end{array}%
\right. $} \\ \hline
\end{tabular}%
\caption{Estimates of the critical and subcritical exponents of the Morris equations (\ref{eq:Morris1}, \ref{eq:Morris2})
at $d=3$ in the LPA, and at the second order of the derivative expansion. On the right of each bracket
are given the results obtained by the mapping (first line) and the shooting (second line) methods used in this work and, 
when they exist,
previous estimates obtained in \cite{3357,3816}  by the shooting method (third line). The estimates of $\omega_2$,  
$\breve{\omega}_{1}$ and $\breve{\omega}_{2}$
are new.
}\label{Tab4}%
\end{table}%

For the determination of the fixed point in the LPA ($d=3$), the maximum
degree of the Taylor polynomial that we have employed is $M=82$. The best
estimation of $k^{\ast },$ given in table \ref{Tab4}, is obtained with $R=2$
and $\alpha =1/2$. The radius of convergence $R_{c}$ of the original series
in powers of $z$ is roughly equal to 2. These facts suggest that the
singularity the closest to the origin lies close to the imaginary axis in
the complex plane of $z$.

The process of determining the eigenvalues is similar to the cases already
described in the preceding sections.

In the LPA, the eigenvalues have been estimated with shorter series than the
fixed point, $M=40$ and $M=60$ for the even and odd cases, respectively,
whereas $R$ and $\alpha $ remained unchanged. The clear distribution of the
eigenvalues shown in figure \ref{fig4} is preserved at the LPA and $O\left(
\partial ^{2}\right) $. This particularity of the mapping method is very
interesting since it is not always easy to clearly determine the spectrum of
the eigenvalues (using the shooting method for example, see table \ref{Tab4}%
). In the LPA the mapping method provides essentially the same accuracy than
the shooting method.

\subsection{Second order}

At the next order $O\left( \partial ^{2}\right) $, the main difference with
LPA is the number of equations: two instead of one. For the sake of
shortness we shall not describe again the different steps of the
calculations.

For the fixed point, the maximum degree of the Taylor polynomial we have
attained is $M=17$. Moreover, for small values of $M$, it has been difficult
to locate the correct solution of the fixed point equation because the
auxiliary polynomial conditions have a large number of real zeros.
Nevertheless we have been able to distinguish the correct zeros. The values
of $k^{\ast }$ and $\eta ^{\ast }$ displayed in table \ref{Tab4} have been
obtained with $R=2.5$ and $\alpha =2$. We have estimated the common radius
of convergence to be $R_{c}\left( z\right) \simeq 2.5$. For other values of $%
R$ and $\alpha $ we have observed different kinds of convergence towards a
unique couple of values. So, we are able to provide a rough estimate of the
error on the evaluation of the $O\left( \partial ^{2}\right) $-fixed point
by the mapping method. For $M=17$, it comes:%
\begin{eqnarray*}
k^{\ast } &=&0.258204\pm 0.000023\,, \\
\eta ^{\ast } &=&0.05388\pm 0.00011\,.
\end{eqnarray*}

Regarding the eigenvalues, the largest degree of the Taylor polynomials
considered is $M=25$. The parameters $R$ and $\alpha $ were not necessarily
fixed to the values mentioned above for the fixed point. The eigenvalues
displayed in table \ref{Tab4} have been obtained with $R=2$ and $\alpha =2$.
In addition to those eigenvalues, we have observed in the even case the
presence of a very small eigenvalue ($\simeq 1.5\times 10^{-6}$). It
corresponds to the expected zero eigenvalue associated with the
reparameterization invariance. Its absolute value may be seen as a rough
measure of the accuracy of the calculations.

It is worth indicating that our $O\left( \partial ^{2}\right) $-results are
better than the estimates:%
\begin{eqnarray*}
\eta &=&0.05425\,, \\
\nu &=&0.617476\,,
\end{eqnarray*}%
obtained in \cite{3553} using an expansion about an adjustable non-zero
value of the field.

\subsection{Remark}

Let us emphasize that table \ref{Tab4} displays for the first time, for the
Morris equations (\ref{eq:Morris1},\ref{eq:Morris2}), values of the
eigenvalues in the odd case ($\breve{\omega}_{1}$, $\breve{\omega}_{2}$)
together with the first estimate of $\omega _{2}$. The exponent $\breve{%
\omega}_{1}$ is of particular interest since it is actually the first time
that it is evaluated at the second order of the derivative expansion of an
ERGE. Moreover, this quantity is not well determined since only two
estimations are available: from the Wilson ERGE \cite{440} using the scaling
field method in \cite{319}:%
\begin{equation*}
\breve{\omega}_{1}=2.4\pm 0.4\,,
\end{equation*}

and from the $\epsilon $-expansion up to the order $\epsilon ^{3}$ in \cite%
{3437} from which we have extracted the value:%
\begin{equation*}
\breve{\omega}_{1}=2.34\pm 0.49\,.
\end{equation*}

These two results are close to each other and clearly compatible. This is
not the case of our present evaluation from the Morris equations (\ref%
{eq:Morris1}, \ref{eq:Morris2}). Indeed, assuming that the derivative
expansion converges, we may roughly give an error bar on the estimate of $%
\breve{\omega}_{1}$. Using the two values of table \ref{Tab4}, it comes:%
\begin{equation*}
\breve{\omega}_{1}=1.34\pm 0.5\,,
\end{equation*}%
a result which is incompatible with the two preceding estimates. It is thus
likely that the present $O\left( \partial ^{2}\right) $-calculations do not
provide the correct value of $\breve{\omega}_{1}$, either because the
derivative expansion itself does not converge or simply because the Morris
equations are not adequate to determine correctly the values of the critical
exponents (at least at low orders of the derivative expansion). We think
that the second possibility is the right explanation. This is because one
has already observed in the results of \cite{3828} a bad convergence on $%
\omega _{1}$ and also on $\eta ^{\ast }$ as functions of the number $N$ of
components of the field in the $O\left( N\right) $-symmetry whereas another
choice of cut-off function, as done in \cite{4754}, seems to be better
adapted (see \cite{6138} for example)$.$ It is very likely that the
fundamental reason why the Morris equation is not efficient must be looked
for in the constraint imposed to the $O\left( \partial ^{2}\right) $%
-equations with a view to satisfy explicitly the reparameterization
invariance. Perhaps this constraint is too strong at this order.

\section{Summary and conclusion}

We have presented a new efficient analytical method for solving two point
boundary value problems when one of the two boundaries is located at
infinity. The method is based on the analyticity of the solution looked for
in an angular sector which contains the positive real axis including the
origin. This angular sector is then conformally mapped onto the unit disc so
that an expansion in powers of the new independent variable $w$\ of the
solution looked for yields a convergent series on the whole disc. The
conditions at the infinite boundary may then be safely imposed on an
approximate solution written as a Taylor polynomial in $w$ estimated when $%
w\rightarrow 1$.

We have illustrated and tested the method on three different kinds of ODEs
that one encounters in the derivative expansion of the ERGE: the
Wegner-Houghton and the Wilson-Polchinski equations in the LPA and, at
second order of the derivative expansion, the Morris equations. For the
first and second examples (LPA) we provide the best estimates of both the
fixed point connection parameter and of the critical exponents ever produced
up to now from the ODEs considered. For the second example we are able to
follow the fixed point solution from the dimension $d=3$ down to $2$ where
the kind of solutions we track disappears. For the intermediate dimensions $%
d=8/3$ and $d=5/2$, we are able to locate the existing multi-critical points
\ and the critical exponents attached to them with great accuracy. The
difficulties increase with the multiplicity of the fixed point and with the
decreasing of $d$. The study of the Morris equations is more demanding since
it involves two ODEs (second order of the derivative expansion) which are
particularly difficult. We show that the method works again well. We are
able to estimate subcritical exponents that were not calculated previously
using the derivative expansion. The estimation of one of these subcritical
exponents does not compare favourably with values coming from the $\epsilon $%
-expansion and the scaling field method for studying the ERGEs. We give some
possible explanations why the two equations considered are presumably not
the best choice for calculating the critical exponents.

Comparisons with other analytical methods show that the mapping method is
actually very efficient. It may be useful in the study of more complicated
systems of equations obtained from the derivative expansion.

Obviously, it can also be employed efficiently to study analogous kinds of
ODEs that appear in other fields when\ one of the boundaries is at infinity.

\section{Acknowledgements}

We thank B. Delamotte for indicating us the difficulty of studying the LPA
using a field expansion when the space dimension approaches the value two.

\end{document}